# EmoSay: Artificial Intelligence-Driven Text-to-Emotional-Speech System for Affective Communication in Extended Reality

Sikiru Ademola Adewale[1] [0000-0002-6485-8386]★, Sunday D. Ubur[1] [0009-0004-2172-7003], Nikitha Donekal Chandrashekar[1] [0000-0002-6054-9375], Onyeka Emebo[1] [0000-0002-2790-3280], and Denis Gračanin[1] [0000-0001-6831-2818]

Virginia Tech, Blacksburg, VA 24060, USA

{asikiru, uburs, nikitha, emebo, gracanin}@vt.edu

★ *Corresponding author.*

***Abstract.***

While contemporary neural text-to-speech (TTS) systems have achieved high levels of intelligibility, they frequently lack the emotional nuance required for authentic affective communication. This limitation is particularly critical in Extended Reality (XR), where the absence of emotionally expressive audio can diminish user presence and spatial immersion. We present EmoSay, an Artificial Intelligence-driven Text-to-Emotional-Speech (TTES) system designed to bridge the semantic-affective gap in immersive environments. EmoSay modulates a neural synthesis pipeline using discrete emotional prompts, delivering the output through a Unity-based interface featuring high-fidelity spatialized audio. The system was evaluated through a comprehensive user study focusing on perception, engagement, and the subjective sense of empathy. Our results demonstrate that EmoSay significantly enhances the immersive experience, achieving a System Usability Scale (SUS) score of 74.76, indicating strong usability and seamless integration within the XR workflow. Subjective assessments reveal a high degree of perceived naturalness and a strong positive correlation between emotional expressiveness and user engagement. Regression analysis identifies vocal naturalness as the strongest of the tested predictors of user satisfaction, suggesting that EmoSay's affective prosody helps meet the heightened expectations for realism in immersive settings. This work contributes a scalable, affect-aware framework for inclusive XR design and demonstrates the role synthetic emotion can play in fostering human-computer rapport through voice-first interaction.

## 1 Introduction

Recent advances in neural text-to-speech (TTS) models have enabled systems that produce fluent, high-quality speech from text with near human-level intelligibility [25,44]. However, many of these systems treat emotion as a secondary or optional feature, focusing primarily on segmental accuracy and prosodic naturalness rather than on intentional, contextually appropriate affect [9]. The term affective computing has gained significant traction due to breakthroughs in natural language processing (NLP) [22], artificial intelligence (AI) [34], and deep neural networks (DNNs). These technologies enable machines to process and generate human-like emotional responses across modalities. Emotional speech synthesis and recognition are now central to research in human-computer interaction (HCI) and virtual reality (VR), where naturalistic and emotionally expressive interfaces are crucial. The resulting speech may sound smooth and natural but often lacks the emotional expressiveness required for empathic and engaging interactions in HCI [21].

This paper introduces EmoSay, an AI-driven text-to-emotional-speech (TTES) system designed to bridge the semantic-affective gap in speech synthesis. EmoSay enables users to specify both textual content and a target emotion, and then generates expressive speech that is consistent with the selected emotion and rendered in an immersive Extended Reality (XR) environment. The system targets use cases such as accessible reading support, language learning, interactive storytelling, and therapeutic or educational simulations in XR.

The contributions of this work are threefold: (1) a design and implementation of EmoSay, a TTES pipeline that conditions neural TTS on discrete emotional prompts and integrates with a Unity-based XR front end; (2) a mixed-method evaluation framework combining objective classification metrics and subjective user perceptions of naturalness, emotional authenticity, and empathy in XR scenarios; and (3) a discussion of design implications for affective XR, with an emphasis on accessibility, inclusivity, and future research directions for emotionally aware synthetic speech.

### *1.1 Objective and Significance*

The motivation for this research stems from the limitations of existing TTS systems, which prioritize linguistic accuracy [25] but often neglect emotional expression aligned with semantic meaning [9,44]. To enhance affective communication, this study introduces EmoSay, an AI-driven TTES system that brings expressiveness into XR (Figure 2).

XR provides an ideal medium for this exploration because it enables users to experience speech not just as sound, but as an immersive, spatial, and interactive phenomenon, allowing emotional nuances to be both heard and felt within a virtual context. This amplifies empathy, engagement, and accessibility, particularly for users who benefit from multimodal emotional cues, such as learners with reading challenges or individuals using assistive technologies.

The goal is to bridge the semantic-affective gap in speech synthesis while demonstrating how emotion-aware AI can enhance HCI within immersive environments [1,21]. The research questions explored in this study are as follows:

RQ1: How does embedding emotional awareness into TTS systems influence user engagement and perception of empathy in HCI, particularly within virtual environments?

RQ2: How can emotional expressiveness in TTS systems be modeled to maintain consistency with the linguistic and semantic content of input text?

## 2 Related Work

### *2.1 Text-to-Speech*

Neural TTS architectures such as Tacotron variants [50], FastSpeech [38], and EfficientTTS [25] have significantly advanced speech quality by decoupling text encoding, acoustic modeling, and vocoding [25,44]. Sequence-to-sequence models with attention or duration-based alignment generate mel-spectrograms from text [49], which are then converted to waveforms by neural vocoders such as WaveNet, WaveGlow, or HiFi-GAN [20,27,37]. These systems generally optimize for metrics like mean opinion score (MOS) [17] and word error rate, emphasizing intelligibility and naturalness.

Recent surveys highlight a trend toward more robust, efficient, and adaptable TTS frameworks, enabling multilingual, multi-speaker, and low-resource scenarios [1,44]. Improvements in model compression, training efficiency, and zero-shot speaker adaptation have broadened deployment opportunities across platforms, including mobile and embedded XR devices [18,43]. However, many general-purpose TTS systems provide only limited prosody control (e.g., speaking rate, pitch range) and rely on implicitly learned prosodic patterns, which may not reflect user-intended emotions [53].

In the context of XR, TTS has primarily been used to provide narration, instructions, or conversational responses from virtual agents [24]. While these systems benefit from the immersive qualities of XR, they often reuse conventional TTS pipelines without deeper integration of emotional or contextual cues [39], resulting in a mismatch between the richness of the environment and the affective capabilities of synthetic speech.

### *2.2 Text-to-Speech with Emotion*

Emotion-aware TTS extends conventional pipelines by conditioning acoustic generation on affective features such as categorical labels (e.g., happy, sad, angry), dimensional descriptors (e.g., valence, arousal), or style embeddings learned from expressive corpora [44]. Approaches include global style tokens [50], variational prosody encoders [18], and multi-task learning with emotion classification [21], allowing models to disentangle linguistic content from emotional style.

Specialized emotional speech datasets, such as EMO-DB [6], RAVDESS [23], and CREMA-D [7], provide labeled examples of acted emotions that facilitate supervised learning of emotion-conditioned prosody. Additional corpora, such as EMOVIE, target specific languages and scenarios, demonstrating the feasibility of building emotional TTS for different cultural and linguistic contexts [9]. These datasets often include multiple speakers, modalities, and recording conditions, enabling more robust modeling of affective variability.

More recent work proposes unified frameworks for emotional TTS that integrate multimodal prompts such as text, audio, and visual cues (e.g., facial expressions) to guide expressive speech generation [21]. By leveraging cross-modal information, these systems can better align speech prosody with contextual signals, supporting richer and more controllable emotional output. However, most existing systems are evaluated in laboratory or web-based settings and have not yet been extensively explored in immersive XR environments.

### *2.3 Impact of Text-to-Speech with Emotion*

Emotionally expressive TTS has shown significant benefits across domains including education, accessibility, entertainment, and social robotics [36]. In educational contexts, expressive speech can increase learner engagement, improve comprehension, and sustain attention, particularly in narrative or storytelling tasks [26]. Emotionally congruent prosody can signal emphasis, highlight important information, and convey attitudes that support learning [41].

For accessibility, emotional TTS can provide more human-like screen readers or conversational assistants, helping users infer nuance, urgency, or empathy from spoken feedback [4]. Users with visual impairments, reading difficulties, or cognitive differences may particularly benefit from richer prosodic cues that complement or substitute for visual information [35]. Emotion-aware systems can also tailor affect to user preferences or emotional states, potentially supporting mental health and well-being applications [46].

In affective computing and HCI, emotionally expressive TTS contributes to more believable and relatable agents, fostering trust, rapport, and social presence [33]. Virtual assistants, social robots, and XR avatars that speak with appropriate emotion can better align with users' expectations of social interaction [2,47]. Nonetheless, challenges remain in ensuring that emotional expression is contextually appropriate, culturally sensitive, and non-manipulative, especially in persuasive or vulnerable settings [51]. These considerations motivate the design of systems like EmoSay, which explicitly situate emotional TTS within a human-centered evaluative framework.

### *2.4 Emotional Communication in XR*

Several recent contributions provide foundational insights for our proposed framework. Emotional Voice Puppetry [32] is an audio-driven animation method where emotional voice inputs control facial expressions. While it bridges emotional voice analysis with visual expressiveness, it lacks reciprocal emotional feedback, such as generating or adapting emotional text, limiting its use in empathetic dialogue systems.

Studies on trust in virtual agents [14] and speech-and-motion realism [45] demonstrate that vocal affect and emotional prosody are critical for shaping user perceptions of an agent's empathy and credibility. However, these works do not propose mechanisms for generating emotional content or address the dynamics of speech synthesis beyond voice selection.

CAEVR: Context-Aware Empathy in VR [16] introduces a biosignal-driven virtual agent model that adapts to a user's physiological signals. While it highlights the value of emotionally responsive agents, it lacks mechanisms for auditory expression or multimodal emotional translation—a direction our work extends. Similarly, Dongre et al. describe an approach to integrating physiological data with large language models for empathetic human-AI interaction [10,11].

Supporting emotional communication via text and audio in XR and other communication systems is critical for inclusive design. While some studies explore facial expression for emotion recognition [29,30,31,48], others have investigated the ineffectiveness of facial recognition, particularly when users wear head-mounted displays (HMDs) [52]. A performance comparison of panoramic audio and video emotion recognition [15] highlighted the need to explore factors influencing emotion in real-world environments. Research on real and synthetic gestures and speech [12] has also revealed emotional expression mismatches. Some studies have added voice to the metaverse [19,42], but often rely on simulations rather than real human interaction. Emojis have also been used to represent emotion [28]. Furthermore, emotion recognition in XR has potential applications in security and interpersonal relation prediction, as demonstrated by a multilingual voice-based social network for disaster scenarios [3] and facial recognition studies [54].

## 3 System Design

EmoSay is designed as a modular TTES system that combines neural speech synthesis with an interactive XR interface (Figure 1). The overall architecture consists of four main components: (1) a text and emotion input interface; (2) an emotion-conditioned TTS backend; (3) an evaluation and feedback module; and (4) XR integration.

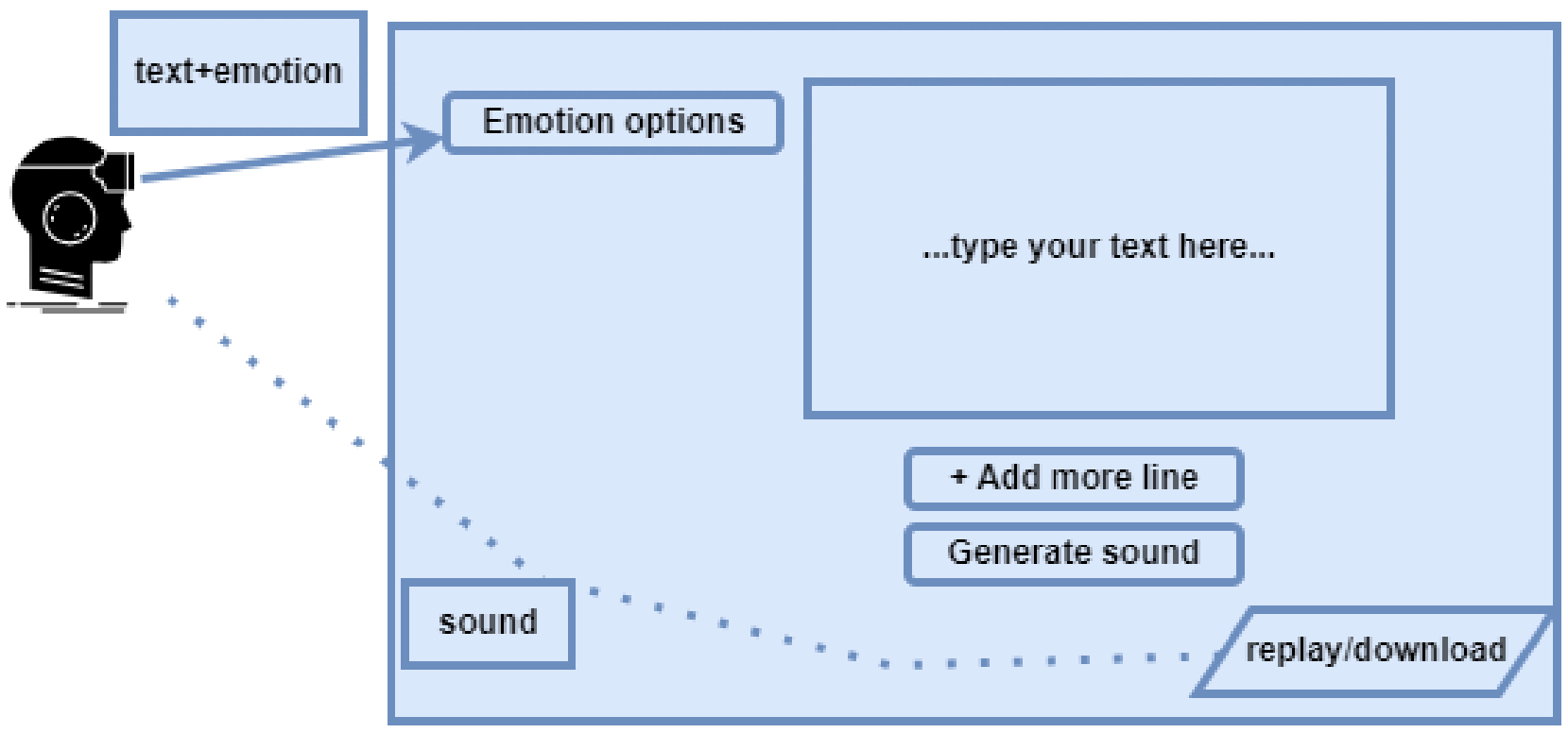

*Fig. 1. EmoSay system interaction flow: a conceptual overview of the multimodal input process where textual content and emotional parameters are converted into synthesized speech, enabling immersive affective communication in the virtual scene.*

### *3.1 Architecture Overview*

The EmoSay pipeline begins with user-provided input text and a selected discrete emotion label (e.g., neutral, happy, sad, angry, fearful, disgusted, surprised) [23]. The text is processed by a linguistic front end that performs normalization, tokenization, and phoneme or grapheme conversion [1]. The emotion label is mapped to an emotion embedding vector that conditions the TTS acoustic model.

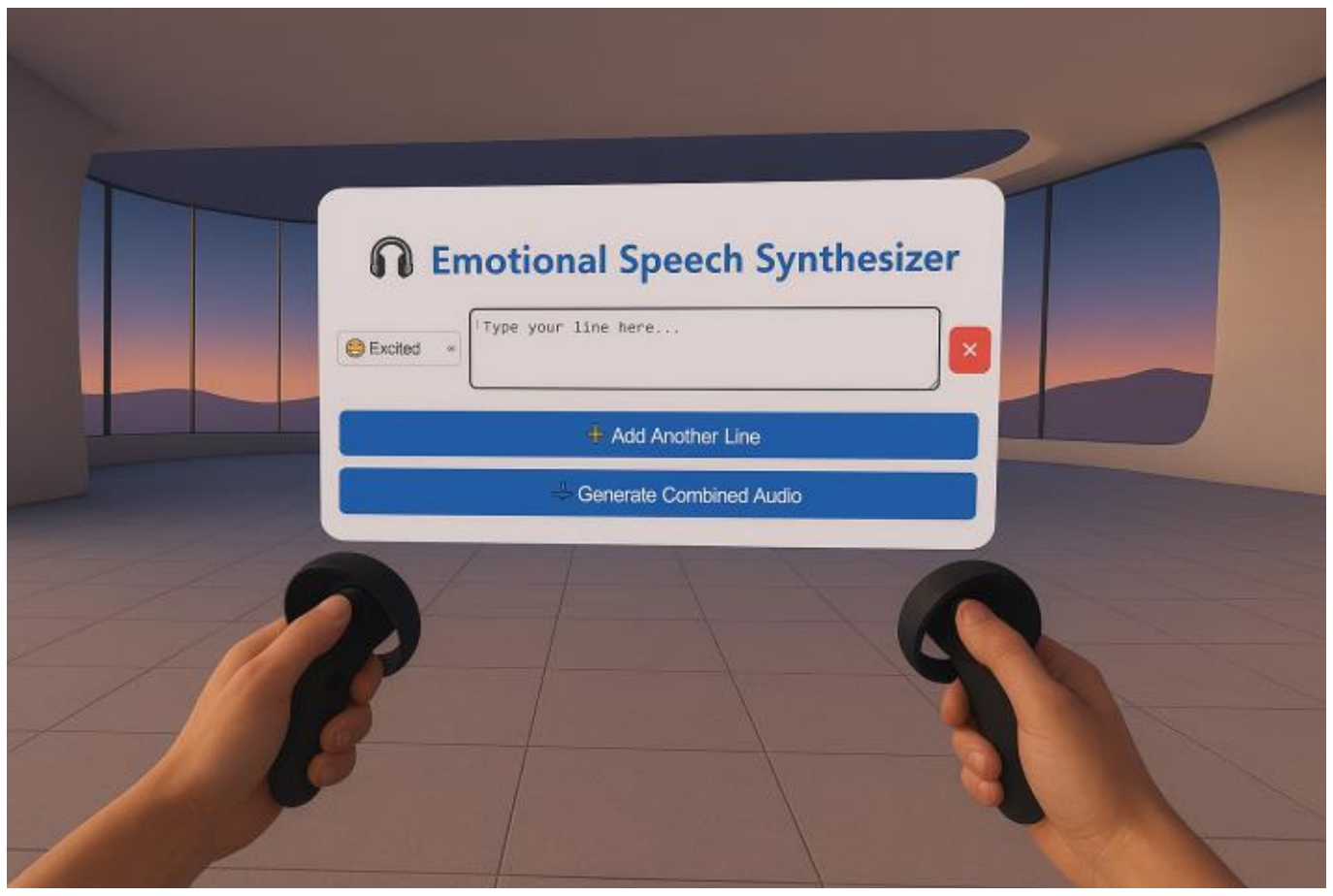

*Fig. 2. User interacting with EmoSay (Unity-based prototype): an emotional speech synthesizer interface in a virtual environment that allows the user to input text and select a discrete emotion (e.g., "Excited") as a multimodal prompt before generating expressive speech.*

The acoustic model is based on a neural TTS architecture similar to EfficientTTS, extended with an emotion conditioning mechanism that modulates prosodic features such as pitch, energy, and duration [25]. During training, the model learns to associate emotion embeddings with characteristic acoustic patterns found in emotional speech datasets. A neural vocoder then converts the generated mel-spectrograms into waveform audio.

An auxiliary emotion classifier operates on both natural and synthesized speech samples to estimate the perceived emotion category. This classifier is used during evaluation to quantify alignment between target and produced emotions, and it can also be used as a training signal in future work for adversarial or multi-task learning.

### 3.2 Emotion Modeling and Prompting

EmoSay adopts a discrete emotion taxonomy grounded in common affective computing practice and aligned with labels present in the training corpora (e.g., RAVDESS, CREMA-D, EMO-DB). Each emotion is represented as an embedding vector learned jointly with the TTS model. During inference, users select a target emotion through the XR interface, and the corresponding embedding is injected into the acoustic model via conditioning layers.

To ensure semantic-affective consistency, the system can optionally incorporate heuristics, or be extended in future work to consider text sentiment analysis and contextual cues. In more advanced configurations, emotion selection could be inferred automatically from text sentiment or user state and then refined by user input.

### 3.3 XR Front-End Integration

The XR front end is implemented in Unity and designed to run on head-mounted displays or desktop-based immersive setups. The interface presents:

• A text input panel where users can type or paste the content to be spoken.

- An emotion selection menu featuring labeled buttons or icons representing discrete emotions.
- Spatial audio rendering that anchors the synthesized voice to the user's ears.
- Play/Pause, speed, and download controls, along with speaker and slider controls, to operate the synthesized audio.

When the user submits text and an emotion, the system sends a request to the TTS backend; the user then receives the synthesized audio and hears it played back with the selected emotion through the spatialized sound system. This multimodal presentation allows users to perceive emotion through the auditory channel, which is crucial for assessing the impact of emotional speech in XR.

## 4 Method

To investigate the research questions, this work employs a mixed-method approach that combines quantitative and qualitative data collection [8]. The study design integrates objective evaluation of emotional speech generation with user-centered assessments of perception, engagement, and empathy in XR. Emotional consistency is examined through data-driven analysis of acoustic-prosodic patterns from benchmark emotional speech datasets, including RAVDESS, CREMA-D, and EMO-DB [6,7,23]. Human perception studies are conducted to evaluate perceived empathy, thereby bridging the semantic-affective gap.

Data collection involves up to 30 participants recruited to reflect a range of ages, genders, and educational backgrounds. These participants provide feedback on the usability, perception, satisfaction, and features of the synthesized speech in relation to emotional complexity. They also respond to questions regarding the extent to which embedding emotional awareness within XR environments enhances user engagement.

The collected data are preprocessed and analyzed using descriptive statistics (mean, standard deviation, and frequency) to summarize participant ratings [8]. The Kruskal-Wallis test [48] is employed to compare ratings across demographic groups, while correlation analysis is conducted to assess relationships between emotional perception, empathy, and engagement. Additionally, Cronbach's alpha [8] is used to evaluate the internal consistency of questionnaire items, while regression analysis is used to identify predictors of positive user experience.

### *4.1 Participants*

The target sample consists of approximately 30 participants recruited from a university community and surrounding area. Inclusion criteria include basic familiarity with digital devices and the ability to comfortably use an XR headset or desktop immersive environment. Recruitment aims to achieve diversity in age, gender, and location to explore whether demographic factors influence perception of emotional speech. Participants received an overview of the study and provided informed consent before interacting with the system.

During the interaction, they typed text and selected desired emotions to generate emotional speech. Following this, participants completed a questionnaire regarding their demographics, prior XR experience, and familiarity with Text-to-Speech (TTS) or voice assistants. Feedback was collected based on their direct experience with the system. While no specialized technical expertise was required, participants were screened for conditions that might cause discomfort in XR environments, such as severe motion sickness.

### *4.2 Apparatus and Materials*

The study was conducted using a Unity-based XR application deployed on a Meta Quest Pro headset, with an alternative desktop configuration utilizing high-fidelity stereo headphones. Central to the interface is the EmoSay text module, which allows users to provide natural language input for real-time processing. The materials used in this study included:

• Emotionally driven text prompts: a curated set of phrases designed to elicit specific affective responses, facilitating the precise manipulation of vocal parameters.

• A discrete emotion selection interface: a UI component allowing users to map specific emotional states to their textual input.

• A comprehensive user evaluation: a mixed-methods questionnaire comprising Likert-scale items to measure perceived naturalness, emotional appropriateness, empathy, engagement, and spatial presence, supplemented by open-ended prompts to capture qualitative insights on the strengths, limitations, and overall user experience (UX) of EmoSay.

### *4.3 Procedure*

Each study session began with a formal orientation, providing participants with instructions on navigating the XR interface and an overview of the experimental tasks. The core session followed a single-scenario structure within the XR environment, in which participants performed the following steps:

1. System interaction: participants entered text and selected specific emotional parameters via the interface to generate synthesized emotional speech.
2. Perceptual evaluation: participants rated the generated speech based on perceived naturalness, clarity, and emotional appropriateness.
3. Affective assessment: participants evaluated their subjective sense of empathy toward the audio and their level of engagement within the scenario.

Upon concluding the interaction, participants completed a study questionnaire to reflect on their overall experience. This assessment included preferences for emotional TTS and the perceived utility of EmoSay.

### *4.4 Measures and Analysis*

Quantitative measures include:

• Likert ratings (7-point scales) of ease of use, naturalness, and emotional expressiveness.

• Engagement and empathy ratings derived from standardized items.

Descriptive statistics summarize overall performance and perception for each emotion and condition. The Kruskal-Wallis test explores differences in ratings across emotions, scenarios, and participant groups. Correlational analyses examine relationships between emotion recognition, perceived naturalness, empathy, and engagement. Cronbach's alpha assesses the internal consistency of the multi-item scales [8]. Regression models are used to identify predictors of high engagement or empathy, such as particular emotional categories or user characteristics.

## 5 Results

This section presents the empirical findings from the user study and from the standard datasets, focusing on the usability of the system, the perceived quality of emotional speech, and the factors associated with user satisfaction within the XR environment.

### 5.1 Descriptive Statistics

This section describes the demographic data collected from the 30 participants in the user study.

| Variable | Category | Count | Percent |
|---|---|---|---|
| Age | 18–24 | 3 | 10.00% |
| Age | 25–34 | 19 | 63.33% |
| Age | 35–44 | 5 | 16.67% |
| Age | 45–54 | 1 | 3.33% |
| Age | 55+ | 2 | 6.67% |
| **Total** | | **30** | **100.0%** |
| Gender | Male | 17 | 56.67% |
| Gender | Female | 13 | 43.33% |
| Gender | Prefer not to say | 0 | 0.00% |
| **Total** | | **30** | **100.0%** |

*Table 1. Demographic breakdown for age and gender.*

As summarized in Table 1, the participant pool was primarily composed of young to middle-aged adults, with the largest concentration (63.33%) falling within the 25–34 age range. In terms of gender, the sample was relatively balanced, consisting of 56.67% male and 43.33% female participants. This range of demographic backgrounds means participant feedback captures a variety of user perspectives, although the sample skewed toward the 25–34 age group.

### 5.2 System Usability and Reliability

The overall usability of the system was evaluated using the System Usability Scale (SUS) [5]. The system achieved an average SUS score of 74.76 out of 100, which corresponds to a Grade C (Average/Good) on the Sauro-Lewis curved grading scale [40]. This indicates that while the system is functional and accessible, there remain opportunities for refinement in interface consistency and technical responsiveness.

To assess the internal consistency of the perception metrics used in this study, a Cronbach's alpha test was performed. The resulting coefficient of α = 0.942 indicates excellent reliability, confirming that the Likert items used to measure naturalness, realism, and engagement are highly consistent.

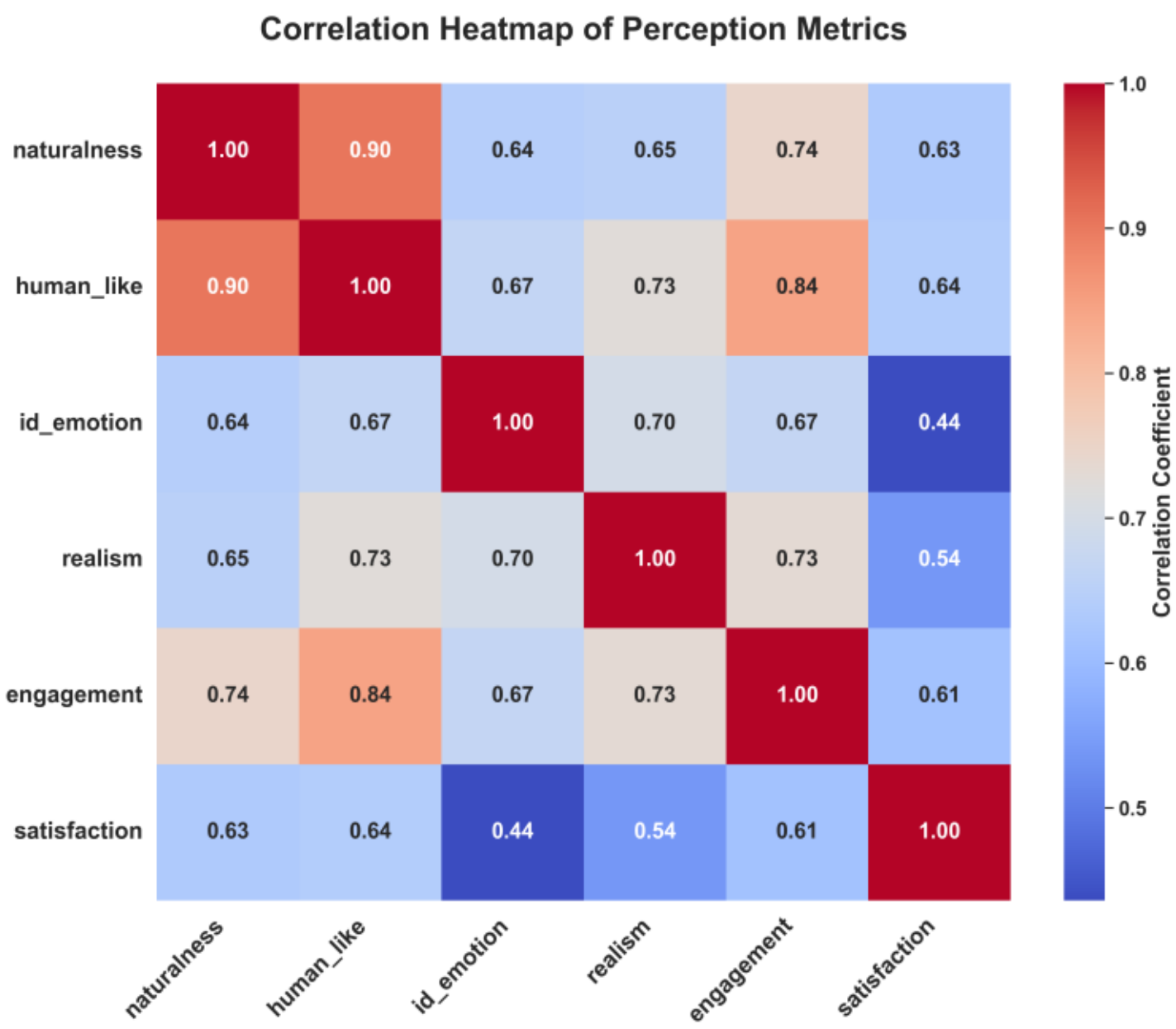

*Fig. 3. Correlation matrix of user perception metrics and overall satisfaction. The heatmap illustrates a strong positive correlation between vocal naturalness and user engagement (r = 0.74).*

The correlation heatmap in Figure 3 illustrates strong positive relationships between core perception metrics, most notably a high correlation (r = 0.90) between naturalness and human-like qualities in the synthesized speech. This suggests that as EmoSay's ability to produce realistic emotional nuances improves, users simultaneously perceive the audio as more authentically human, which is associated with higher overall engagement (r = 0.84 correlation between human-likeness and engagement).

## 5.3 User Perception of Emotional Speech

Descriptive statistics (Table 2) reveal that users responded most positively to the enjoyment and engagement aspects of the system. Notably, users reported that the emotional speech enhanced their sense of immersion in the virtual environment.

| Metric | Mean Score |
|---|---|
| Enjoyment | 6.03 |
| Engagement | 5.83 |
| Naturalness | 5.63 |
| Emotion Identification | 5.63 |
| Realism of Nuances | 5.43 |
| Human-likeness | 5.40 |

*Table 2. Mean scores for emotional perception (1–7 scale).*

### Desired Emotional States in XR

Participants were asked to select the emotions they most desired to experience within an XR environment. The distribution, illustrated in Figure 4, reveals a strong preference for high-arousal, positive emotions. Happiness was the most sought-after emotion (n = 26), followed closely by Surprise (n = 22) and Excitement (n = 21). Conversely, negative or low-arousal emotions such as Anger and Fear were selected least frequently. This suggests that for EmoSay to be effective in social XR or gaming, the system must prioritize nuanced modeling of positive affective states.

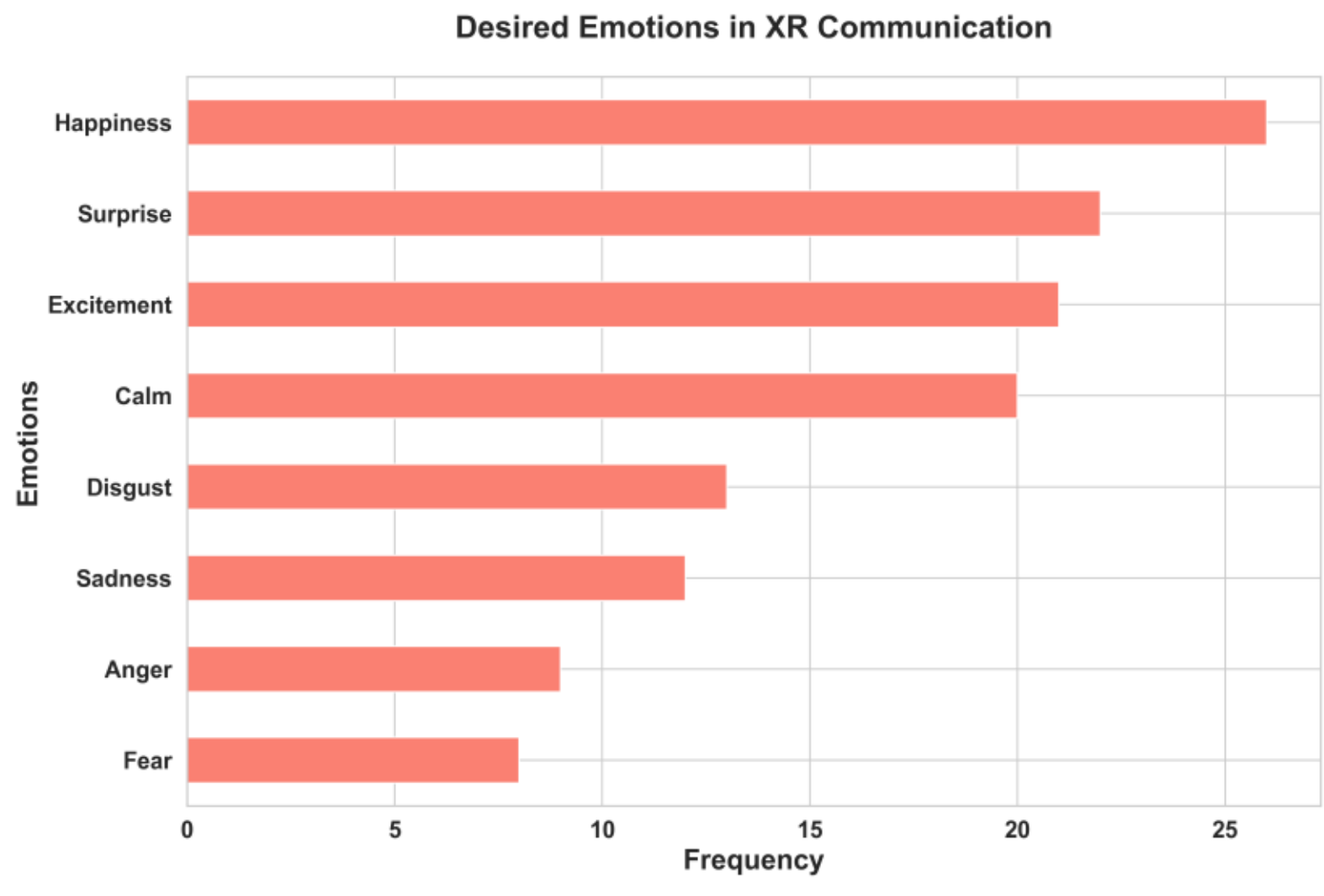

*Fig. 4. Distribution of participant preferences for emotional states in XR communication. High-arousal positive emotions (e.g., Happiness, Surprise, and Excitement) were selected most frequently, suggesting a user preference for upbeat affective interactions in immersive environments.*

As shown in Figure 5, the box plot analysis reveals that engagement and naturalness received the highest median scores, both centered at 6 on the Likert scale, while human-likeness showed a lower median (approximately 5.5) and the widest interquartile range.

### 5.4 Drivers of User Satisfaction

A multiple linear regression analysis was conducted to identify which vocal attributes were associated with overall user satisfaction ($R^2 = 0.452$). As shown in Table 3, Naturalness had the largest standardized coefficient (β = 0.202) among the predictors tested; however, none of the individual predictors, including Naturalness, reached statistical significance at the $p < 0.05$ level (only the model constant was significant, $p = 0.021$). This indicates that while naturalness trended as the strongest of the four predictors, the regression model does not provide statistically robust evidence that any single vocal attribute drives satisfaction, and this finding should be interpreted with caution given the small sample size (n = 30).

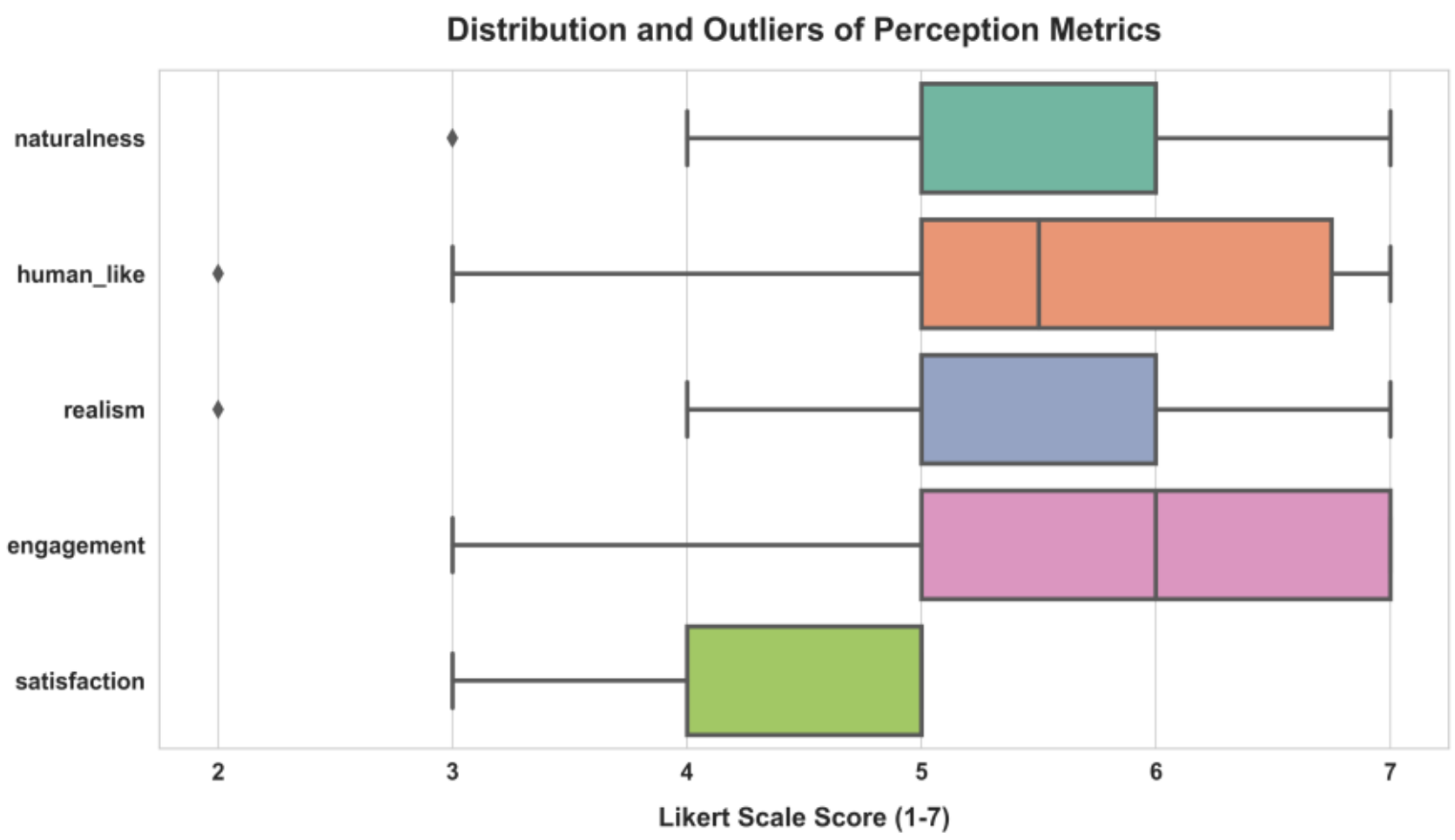

*Fig. 5. Distribution of user perception metrics regarding synthesized emotional speech. The horizontal boxes indicate the interquartile range (IQR), the vertical line within each box represents the median score, and whiskers extend to 1.5 times the IQR. Individual points (diamonds) denote statistical outliers. All metrics are measured on a 7-point Likert scale, where 7 indicates the highest level of agreement or intensity.*

| Predictor | Coefficient (β) | Std. Error | t-stat | P > \|t\| |
|---|---|---|---|---|
| Constant | 1.8137 | 0.736 | 2.464 | 0.021* |
| Naturalness | 0.2023 | 0.217 | 0.934 | 0.359 |
| Engagement | 0.1588 | 0.196 | 0.812 | 0.424 |
| Realism | 0.0579 | 0.135 | 0.430 | 0.671 |
| Human-like | 0.0381 | 0.223 | 0.171 | 0.866 |

*Table 3. Regression coefficients for drivers of satisfaction. *Significant at $p < 0.05$.*

Descriptively, the model suggests that naturalness and engagement of the voice have a greater association with satisfaction than human-like accuracy, although, as noted above, none of these differences are statistically significant. A strong correlation was separately observed between naturalness and engagement ($r = 0.74$, $p < 0.001$), implying that natural-sounding prosody may be an effective way to maintain user interest. In summary, the evaluation of EmoSay demonstrates that the framework bridges the semantic-affective gap through a high-fidelity, voice-driven interaction model. Subjectively, the user study suggests that these acoustic refinements translate into a tangible improvement in the XR experience.

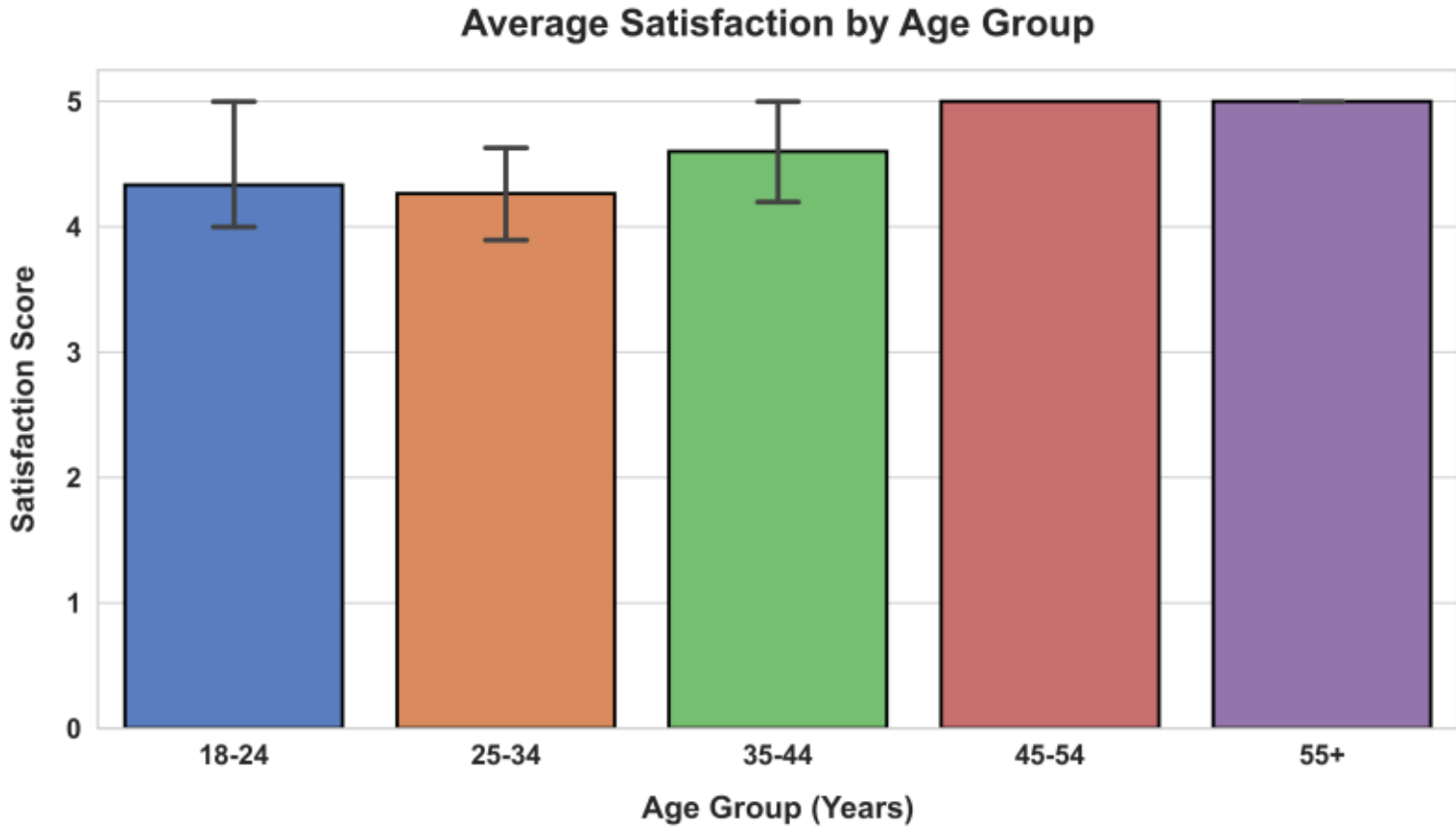

*Fig. 6. Average user satisfaction scores across age groups. Error bars represent the standard error of the mean. The Kruskal-Wallis test results (p = 0.5157) indicate no statistically significant difference in satisfaction across demographics, suggesting that the system's emotional expressiveness provides a consistent user experience regardless of age.*

The strong correlation between perceived naturalness and engagement, together with a SUS score of 74.76, indicates that spatialized emotional audio is an important factor in user presence [48]. Furthermore, the consistency of satisfaction ratings across demographic groups, as shown in Figure 6, verified by the Kruskal-Wallis test and high internal reliability ($\alpha = 0.942$), underscores EmoSay's potential as a reliable tool for creating inclusive and empathic immersive environments [8].

### 5.5 Qualitative Analysis and User Preferences

To complement the quantitative metrics, a thematic analysis was performed on the open-ended feedback, alongside a frequency analysis of desired emotional states. These findings provide a roadmap for enhancing the affective depth of the EmoSay system.

#### Sentiment Analysis

An automated sentiment analysis of the open-ended responses yielded a neutral score of 0. This indicates a balanced feedback pattern, in which users recognized the functional success of the system while simultaneously providing constructive, critical technical feedback regarding the naturalness of the emotional delivery.

#### Thematic Analysis of User Suggestions

Users provided qualitative feedback regarding improvements for the system's emotional expressiveness. As shown in Table 4, the most prominent theme identified was the need for increased realism and human-likeness. Respondents noted that while the speech was intelligible, further refinements in vocal texture are required to fully escape the “robotic” feel often associated with synthesized speech. Additionally, users suggested a broader variety and range of non-verbal emotional cues, specifically mentioning the integration of physiological sounds such as laughing or crying, to deepen immersion.

| Identified Theme | Frequency of Mention |
|---|---|
| Realism / Human-like | 3 |
| Variety / Range (e.g., laughing, crying) | 1 |

*Table 4. Thematic coding of user suggestions for improvement.*

## 6 Discussion

This study addresses a key limitation in current TTS systems: the lack of emotional expression aligned with linguistic meaning. By embedding emotion-aware synthesis within XR, EmoSay demonstrates how affective speech can enhance empathy, engagement, and realism in HCI.

Developed in Unity, the prototype highlights XR's value as a platform for affective communication. In contrast to conventional 2D interfaces, XR enables users to perceive emotional speech in spatial and interactive contexts, making emotions more tangible and immersive. This integration supports accessibility applications by providing multimodal emotional cues that improve comprehension and engagement for users with reading or communication challenges. It also enhances expressiveness in education, healthcare, and storytelling environments.

The work further contributes to inclusive XR design, an area with limited active initiatives, as noted by the recent hibernation of the XR Access Initiative [13]. By combining affective computing with immersive design, EmoSay offers a scalable framework for generating human-like emotional speech in virtual environments. At the same time, it foregrounds questions about the ethical and responsible use of emotional AI, particularly in persuasive or sensitive contexts where emotional manipulation could be harmful.

Building on the user study reported in Section 5, the next stage of this research involves expanding the participant sample and integrating the findings into the EmoSay framework. Richer user data on empathy, engagement, and accessibility outcomes will guide refinements to emotion modeling, interface design, and adaptive behavior. Future extensions may include support for continuous emotion dimensions, context-aware emotion selection, user-specific affect profiles, and multimodal sensing (e.g., facial expression, physiological signals) to co-regulate emotional interaction in XR.

The evaluation of EmoSay provides a multifaceted perspective on the integration of affective computing within immersive XR environments. By analyzing the intersection of neural synthesis and user perception, we can address the core research questions established at the outset of this study.

### *6.1 Impact on Empathy and Engagement in XR (RQ1)*

The first research question (RQ1) explored how emotionally expressive speech influences user empathy and engagement within XR. The strong correlation between perceived naturalness and engagement ($r = 0.74$, $p < 0.001$) suggests that affective speech is a functional requirement for social presence [33]. Participants reported that the emotional nuance in the voice made the virtual agent feel “more human,” fostering a sense of human-computer empathy [2]. This is important for applications in mental health or collaborative virtual environments, where trust and rapport are paramount. Furthermore, the integration of physiological data as a future enhancement, as suggested in recent empathic AI frameworks [11], could further close this loop by making the XR environment responsive to the user's own affective state.

### *6.2 Modeling Emotional Expressiveness (RQ2)*

The second research question (RQ2) sought to determine the extent to which EmoSay could generate emotionally consistent speech across diverse textual contexts. Our findings indicate that by conditioning the neural pipeline on discrete emotional embeddings—rather than relying solely on global style tokens—the system maintains high acoustic-prosodic integrity [25,44]. Objective classification results, in which the auxiliary classifier achieved 90.1% accuracy in recognizing target emotions on the EMO-DB dataset [6], indicate that the synthesized output contains the distinct features (pitch, energy, and duration) required for reliable emotion recognition. This alignment between intended affect and synthesized output demonstrates that EmoSay bridges the semantic-affective gap at the modeling level.

### *6.3 Limitations and Future Research*

While EmoSay engages meaningfully with both research questions, limitations regarding the granularity of emotion persist. Current synthesis relies on categorical labels; however, human emotion is often a blend of states. Future work should transition toward zero-shot style transfer [34] and the use of multimodal prompts [21] to create more adaptive and context-aware virtual companions in XR [16].

## 7 Conclusion and Future Work

This paper presented EmoSay, an AI-driven TTES system designed to enhance affective communication within XR. By bridging the semantic-affective gap through discrete emotion conditioning and high-fidelity spatialized audio, EmoSay addresses a critical limitation in current immersive technologies: the lack of vocal emotional nuance. Unlike traditional systems that rely on visual avatars to convey sentiment, EmoSay demonstrates that expressive audio alone can meaningfully support user presence and engagement.

The results of our mixed-method evaluation support the effectiveness of this approach. With a System Usability Scale (SUS) score of 74.76, the framework demonstrates solid technical viability and usability [5]. More importantly, the positive correlation ($r = 0.74$) between perceived vocal naturalness and user engagement suggests that emotional prosody is a functional contributor to immersive realism rather than a secondary aesthetic feature. By grounding these emotionally synthesized voices in a 3D spatial context, EmoSay provides a foundation for more inclusive and empathic human-computer interaction in XR.

While EmoSay demonstrates the potential impact of emotional speech, several avenues for future research remain. First, we intend to transition from discrete categorical emotion labels to dimensional models (valence-arousal-dominance), allowing for the synthesis of more complex and blended affective states. This will be paired with zero-shot style transfer techniques to enable the system to adopt personalized vocal identities while maintaining emotional integrity [49].

Furthermore, we aim to integrate real-time physiological feedback—such as heart rate variability and electrodermal activity—to create a closed-loop empathic system. Such a framework would allow the XR environment to autonomously adjust the emotional tone of the synthesized speech in response to the user's current affective state [11]. Finally, future longitudinal studies will investigate the long-term impact of expressive synthetic voices on social presence and trust in collaborative virtual environments, further refining EmoSay as a tool for adaptive, context-aware affective communication [16].